\documentclass[twoside,12pt]{article}
\usepackage{epsfig}

\newcommand{\be}{\begin{equation}}
\newcommand{\ee}{\end{equation}}
\newcommand{\bea}{\begin{eqnarray}}
\newcommand{\eea}{\end{eqnarray}}

\begin{document}

\title{ \vspace{1cm} Two, three, many body systems involving mesons}
\author{E.\ Oset$^1$, A.\ Martinez Torres$^2$, K. P.\ Khemchandani$^3$, L.\ Roca$^4$, J.\ Yamagata$^1$\\
$^1$Instituto de F{\'\i}sica Corpuscular (centro mixto CSIC-UV)\\
Institutos de Investigaci\'on de Paterna, Aptdo. 22085, 46071, Valencia, Spain\\
$^2$Yukawa Institute for Theoretical Physics, 
Kyoto University, Kyoto 606-8502, Japan\\
$^3$Research Center for Nuclear Physics (RCNP), Osaka University,\\
 Ibaraki, Osaka 567-0047, Japan\\
$^4$Departamento de Fisica. Universidad de Murcia. E-30071, Murcia. Spain
}

\maketitle

\begin{abstract}
In this talk we show recent developments on few body systems involving mesons.
 We report on an approach to Faddeev equations using chiral unitary dynamics, where an explicit cancellation of the two body off shell amplitude with three body forces stemming from the same chiral Lagrangians takes place. This removal of the unphysical off shell part of the amplitudes is most welcome and renders the approach unambiguous, showing that only on shell two body amplitudes need to be used. Within this approach, systems of two mesons and one baryon are studied, reproducing properties of the low lying $1/2^+$ states. On the other hand we also report on multirho and $K^*$ multirho states which can be associated to known meson resonances of high spin. 
\end{abstract}
%\eject
%\tableofcontents
\section{Introduction}

Faddeev equations \cite{Faddeev,Alt:1967fx} have been and continue to be the standard way to study three body systems. One intrinsic problem, common to most many body approaches, is the dependence of the results on the off shell extrapolation of the two body amplitudes, which is an unphysical magnitude. Removing the ambiguity derived from this unphysical part should be most welcome, but it is generally not possible when one deals with potentials. The advent of chiral unitary dynamics to deal with hadron interactions 
\cite{Kaiser:1995eg,angels,ollerulf,Jido:2003cb,Oset:2001cn,Hyodo:2002pk,Borasoy:2005ie,Oller:2006jw,Borasoy:2006sr,Lutz:2001yb,Nieves:1999bx}  has brough a solution to this problem. Indeed, in recent studies, the use of chiral dynamics in the Faddeev equations has shown that the off shell part of the two body amplitudes, which appears in the Faddeev equations, gets cancelled by three body contact terms stemming from the same chiral Lagrangians \cite{MartinezTorres:2007sr,MartinezTorres:2008gy,Torres:2011jt,Khemchandani:2008rk}. In this way the unphysical part of the amplitudes is removed and only the on shell two body amplitudes are needed in this approach. These off shell effects are responsible for the differences in the three body  calculations that use  input potentials producing the same on shell two body amplitudes. The new method removes these ambiguities and relies upon physical on shell amplitudes.
  We show results for systems of two mesons and one baryon leading to low lying $J^P= 1/2^+$ states.
 The neat reproduction of the low lying $1/2^-$ states in the $S$-wave meson-baryon interaction, using chiral dynamics, suggests that the addition of a pseudoscalar meson in S-wave could lead to an important component of the structure of the $1/2^+$ resonances.  Chiral dynamics has been used earlier in the context of the three nucleon problems, e.g., in \cite{epelbaum}. We present here the study done in \cite{MartinezTorres:2007sr} of two meson - one baryon systems, where chiral dynamics is applied to solve the Faddeev equations. As described below, our calculations for the total strangeness $S$ = -1 reveal peaks in the  $\pi \bar{K} N$ system and its coupled channels which can be identified with the resonances $\Sigma(1770)$, $\Sigma(1660)$, $\Sigma(1620)$, $\Sigma(1560)$, $\Lambda(1810)$ and $\Lambda(1600)$. With the simplication of the Faddeev equation from the Fixed Center Approximation (FCA), we also address the interaction of several mesons and find that multirho and $K^*$ multirho states are obtained and can be associated to mesons of high spin already known and reported in the PDG.

\section{The formalism for three body systems}
We start by taking all combinations of a pseudoscalar meson of the $0^-$ SU(3) octet and a baryon of the $1/2^+$ octet which couple to $S=-1$ with any charge. For some quantum numbers, the interaction of this two body system is strongly attractive and responsible for the generation of the two $\Lambda(1405)$ states \cite{Jido:2003cb} and other $S$ = -1 resonances. We shall assume that this two body system formed by $\bar{K}N$ and coupled channels remains highly correlated when a third particle is added, in the present case a pion. Yet, the formalism allows for excitation of this cluster in intermediate steps. Altogether, we get twenty-two coupled channels for the net charge zero configuration: $\pi^0 K^- p$, $\pi^0\bar{K}^0 n$, $\pi^0\pi^0\Sigma^0$, $\pi^0\pi^+\Sigma^-$, $\pi^0\pi^-\Sigma^+$, $\pi^0\pi^0\Lambda$, $\pi^0\eta\Sigma^0$, $\pi^0\eta\Lambda$, $\pi^0 K^+\Xi^-$, $\pi^0 K^0\Xi^0$, $\pi^+ K^- n$, $\pi^+\pi^0\Sigma^-$, $\pi^+\pi^-\Sigma^0$, $\pi^+\pi^-\Lambda$, $\pi^+\eta\Sigma^-$, $\pi^+ K^0\Xi^-$, $\pi^-\bar{K}^0 p$, $\pi^-\pi^0\Sigma^+$, $\pi^-\pi^+\Sigma^0$, $\pi^-\pi^+\Lambda$, 
$\pi^-\eta\Sigma^+$, $\pi^- K^+ \Xi^0$. We assume the subsystem of particles 2 and 3 to have a certain invariant mass, $\sqrt{s_{23}}$, and the three body $T$-matrix is evaluated as a function of  this mass and the total energy of the three body system. At the end we look for the value of $|T|^2$ as a function of these two variables and find peaks at certain values of these two variables, which indicate the mass of the resonances and how a pair of particles is correlated.  

The input required to solve the Faddeev equations consists of the two body $t$-matrices for the meson-meson and meson-baryon interactions which are calculated from the lowest order chiral Lagrangian following
\cite{npa,angels,Oset:2001cn,Inoue} and using the dimensional regularization of the loops as done in
\cite{ollerulf,Oset:2001cn}, where a good reproduction of scattering amplitudes and resonance properties for the low lying $1/2^-$ states was found. Improvements introducing higher order Lagrangians have been done recently \cite{Oller:2006jw,Borasoy:2005ie,Borasoy:2006sr}, including a theoretical error analysis in \cite{Borasoy:2006sr} which allows one to see that the results with the lowest order Lagrangian fit perfectly within the theoretical allowed bands.

A shared feature of the recent unitary chiral dynamical calculations is the on-shell factorization of the potential and the $t$-matrix in the Bethe-Salpeter equation \cite{angels,ollerulf,Nieves:1999bx,Garcia-Recio:2003ks,Hyodo:2002pk,Borasoy:2005ie,npa}, which is
justified by the use of the N/D method and dispersion relations \cite{nsd,ollerulf}. Alternatively, one can see that the off-shell contributions can be reabsorbed into renormalization of the lower order terms \cite{npa,angels}. We develop here a similar approach for the Faddeev equations.% which results into the on-shell factorization of the two body $t$-matrices, while the necessary off-shell dependence of the propagators is maintained.

The full three-body $T$-matrix can be written as a sum of the auxiliary $T$-matrices $T^1$, $T^2$ and $T^3$ \cite{Faddeev}
\begin{equation}
T=T^1+T^2+T^3
\end{equation}
where $T^i$, $i=1$, $2$, $3$, are the normal Faddeev partitions, which include all the possible interactions contributing to the three-body $T$-matrix with the particle $i$ being a spectator in the last interaction.
The Faddeev partitions satisfy the equations
\begin{equation}\label{eq:Tiorig}
T^i=t^i\delta^3(\vec{k}^{\,\prime}_i-\vec{k}_i)+ t^i g^{ij}T^j + t^i g^{ik}T^k ,
\end{equation}
where $\vec{k}_i$ ($\vec{k}^{\,\prime}_i$) is the initial (final) momentum of the ith particle in the global center of mass system, $t^i$ is the two-body $t$-matrix for the interaction of the 
pair $(jk)$ and $g^{ij}$ is the three-body propagator or Green's function, with $j \neq k \neq i$ = 1, 2, 3

\begin{figure}[ht]
\includegraphics[width=0.5\textwidth] {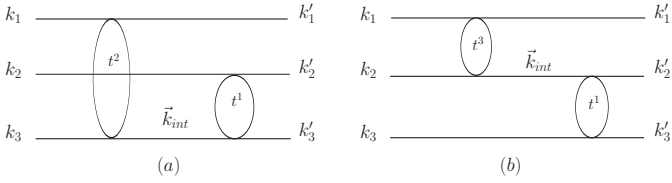}
\caption{\label{fig1} Diagrammatic representation of the terms (a) $t^1 g^{12} t^2$
(b) $t^1 g^{(13)} t^3$.}
\end{figure}

\begin{figure}[ht]
\includegraphics[scale=0.6] {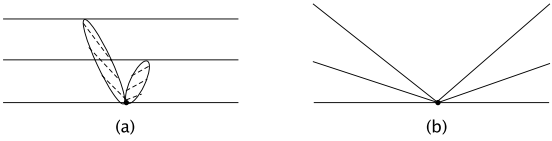}
\caption{\label{fig2} The origin of the three body forces (a) due to cancellation of the propagator in Fig.\ref{fig1}(a) with the off-shell part of the chiral amplitude, (b) at the tree level from the chiral Lagrangian.}
\end{figure}

The first two terms of the diagrammatic expansion of the Faddeev equations, for the case $i$=1, are represented in Fig.\ref{fig1},
where the $t$-matrices are required to be off-shell. However, the chiral amplitudes, which we use,
can be split into an ``on-shell'' part (obtained when the only propagating particle of the diagrams, labeled with $\vec{k}^2_{int}$ in Fig.\ref{fig1}, is placed on-shell (meaning that $\vec{k}^2_{int}$ is replaced by $m^2$ in the amplitudes),  and an off-shell part proportional to the inverse of the propagator of the off-shell particle, $\vec{k}^2_{int}-m^2$. This term would cancel the particle propagator, ($\vec{k}^2_{int}-m^2)^{-1}$, for example that of the 3rd particle in Fig.\ref{fig1}a) resulting into a three body force (Fig.\ref{fig2}a). In addition to this,
three body forces also stem directly from the chiral Lagrangians \cite{felipe} (Fig.\ref{fig2}b).

We find that the sum of the off-shell parts of all the two body interaction terms of the Faddeev series, cancel together with the contribution from Fig.\ref{fig2}(b) in the SU(3) limit. Details of the analytical proof can be seen in the appendices of \cite{MartinezTorres:2007rz,MartinezTorres:2008gy}. Hence, only the on-shell part of the two body (chiral) $t$-matrices is needed in the evaluations. This is one of the important findings of these works because one of the standing problems of the Faddeev equations is that the use of different potentials which give rise to the same on shell scattering amplitudes give rise to different results when used to study three body systems with the Faddeev equations. The different, unphysical, off shell amplitudes of the different potentials are responsible for it. The use of chiral dynamics in the context of the Faddeev equations has then served to show that the results do not depend on these unphysical amplitudes and only the on shell amplitudes are needed as input. In this sense, since these amplitudes can be obtained from experiment, it is suggested in \cite{MartinezTorres:2008kh} to use these experimental amplitudes, and sensible results are obtained in the study of the $\pi \pi N $ system and coupled channels.

 The strategy followed in the former works is that the terms with two, three, and four interactions are evaluated exactly. Then it is observed that the ratio of the four to three body interaction terms is about the same as that of the three body to two body. Once this is realized, the coupled integral equations are converted into algebraic equations, which renders the technical work feasible in spite of the many coupled channels used.

The resonances generated for these system appear as peak in $|T|^2$ as a function of $\sqrt{s}$, $\sqrt{s_{23}}$.
A detailed description of all the states that appear in this sector can be seen in \cite{MartinezTorres:2007sr}. Here we summarize the results in Table \ref{table1}.

\begin{table}
\centering
\begin{tabular}{cccc}
\hline
&$\Gamma$ (PDG)&Peak position& $\Gamma$ (this work)\\
&(MeV)&(this work, MeV)& (MeV)\\
\hline
Isospin=1&&&\\
\hline
$\Sigma(1560)$&10-100&1590&70\\
$\Sigma(1620)$&10-100&1630&39\\
$\Sigma(1660)$&40-200&1656&30\\
$\Sigma(1770)$&60-100&1790&24\\
\hline
Isospin=0&&&\\
\hline
$\Lambda(1600)$&50-250&1568,1700&60-136\\
$\Lambda(1810)$&50-250&1740&20\\
\hline

\end{tabular}
\caption{$\Sigma$ and $\Lambda$ states obtained from the interaction of two mesons and one baryon.}\label{table1}
\end{table}

  In the S=0 sector we also find several resonances, which are summarized in Table \ref{table2}. Here we only want to pay attention to the $N^*$ state around 1924 MeV, which is mostly $N K \bar{K}$. This state was first predicted in \cite{Jido:2008kp} using variational methods and corroborated in \cite{MartinezTorres:2008kh} using coupled channels Faddeev equations. As in \cite{Jido:2008kp}, we find that the $ K \bar{K}$ pair is built mostly around the $f_0(980)$, but it also has a similar strength around the $a_0(980)$, both of which appear basically as a $ K \bar{K}$ molecule in the chiral unitary approach.
  
\begin{table}
\centering
\begin{tabular}{c|ccc|ccc}
\hline\hline
$I(J^P)$&\multicolumn{3}{c}{Theory}&\multicolumn{3}{c}{PDG data}\\
\hline
&channels&mass&width&name&mass&width\\
&&(MeV)&(MeV)&&(MeV)&(MeV)\\
\hline
$1/2(1/2^+)$&only $\pi\pi N$&1704&375&$N^*(1710)$&1680-1740&90-500\\
& $\pi\pi N$, $\pi K\Sigma$, $\pi K\Lambda$, $\pi\eta N$&$\sim$ no change&$\sim$ no change&&&\\
\hline
$1/2(1/2^+)$&only $\pi\pi N$&2100&250&$N^*(2100)$&1885-2270&80-400\\
& $\pi\pi N$, $\pi K\Sigma$, $\pi K\Lambda$, $\pi\eta N$&2080&54&&&\\
\hline
$3/2(1/2^+)$&$\pi\pi N$, $\pi K\Sigma$, $\pi K\Lambda$, $\pi\eta N$&2126&42&$\Delta(1910)$&1870-2152&190-270\\
\hline
$1/2(1/2^+)$&$N\pi\pi $, $N\pi\eta $, $NK\bar K$&1924&20&$N^*(?)$&?&?\\
\hline\hline
\end{tabular}
\caption{$N^{*}$ and $\Delta$ states obtained from the interaction of two mesons and one baryon.}\label{table2}
\end{table}
  
  This state is very interesting and it was suggested in  \cite{conulf} that it could be responsible for the peak around 1920 MeV of the $\gamma p \to K^+ \Lambda$ reaction \cite{saphir,jefflab,mizuki}. It was also shown that the spin of the resonance could be found performing polarization measurements which are the state of the art presently.

  \section{Clusters with many mesons instead of baryons}

    One may wonder why the known nuclei are made of baryons and not of mesons. Certainly, it is not that the interaction between mesons is weaker than between baryons. The reason lies in the property of baryon number conservation which does not hold for mesons. As a consequence, an aggregate of protons and neutrons sufficiently bound has nowhere to decay if baryon number is conserved. However, and aggregate of mesons would decay into systems of smaller number of mesons. Let us then accept that these systems will be unstable, but, even then, could we see them as resonances with a certain width, as most of the particles in the PDG? Gradually an answer is coming to this question. One step in this direction  was given in the study of the three body system $\phi K \bar{K}$ which leads to the recently discovered state $\phi(2170)$, as has been shown in \cite{MartinezTorres:2008gy}. This system, studied through Faddeev equations in coupled channels, produces a state in which the $K \bar{K}$ pair clusterizes into the $f_0(980)$ resonance and the $\phi$ interacts with this cluster. One could proceed further and investigate more complex meson systems. This is what has been recently done in \cite{multirho} where the $f_2(1270)$, $\rho_3(1690)$,   $f_4(2050)$,
 $\rho_5(2350)$ and   $f_6(2510)$ resonances have been described as multi-$\rho(770)$ states.
 
  The idea behind the work of \cite{multirho} is that recent studies show that the $\rho \rho$ interaction is very strong \cite{Molina:2008jw,Geng:2008gx}, particularly when the two $\rho$ mesons align their spins to form a state of spin S=2.
  This interaction is so strong that can bind the two $\rho$ mesons leading to a bound state which, according to \cite{Molina:2008jw,Geng:2008gx}, is the $f_2(1270)$.
 It is surprising to come out with this idea when it has been given for granted that this state and other partners accommodate easily as $q \bar{q}$ states and can reproduce most of the known properties of these states 
\cite{Fariborz:2006xq,Umekawa:2004js,Anisovich:2001zp}. However, it has been shown that with this molecular picture one can reproduce the radiative decay into $\gamma \gamma$ \cite{junko}, the decay of $J/\Psi$
into $\omega (\phi)$ and $f_2(1270)$ (together with other resonances
generated in \cite{Geng:2008gx})  \cite{daizou}, and $J/\Psi$ into $\gamma$
and  $f_2(1270)$ (and the other resonances of \cite{Geng:2008gx})
\cite{hanhart}.  

   Once this is accepted, the idea is to study  a system with three $\rho$ mesons. To give the maximum probability of binding we choose them with their spins aligned to give a state of S=3.  The interaction is studied using the Fixed Center Approximation (FCA) to the Faddeev equations. One obtains the scattering matrix and looks for peaks in $|T|^2$, from where one obtains the mass and the width of the states. One finds a peak that we associate to the  $\rho_3(1690)$. Once this is done then one takes two clusters of $f_2(1270)$ and studies their interaction using as input the scattering matrix $\rho ~ f_2(1270)$ obtained before. The peak obtained in $|T|^2$ can be associated to the    $f_4(2050)$. One further step uses the FCA to study the interaction of a $\rho$ with the $f_4(2050)$, using as input the $\rho ~ f_2(1270)$ interaction obtained before. In this way one obtains a peak that is associated to the  
   $\rho_5(2350)$. A further step, letting an $f_2(1270)$ interact with the 
  $f_4(2050)$ previously obtained, is done and then a peak in  $|T|^2$ appears which we associate to the  $f_6(2510)$.

\begin{figure}
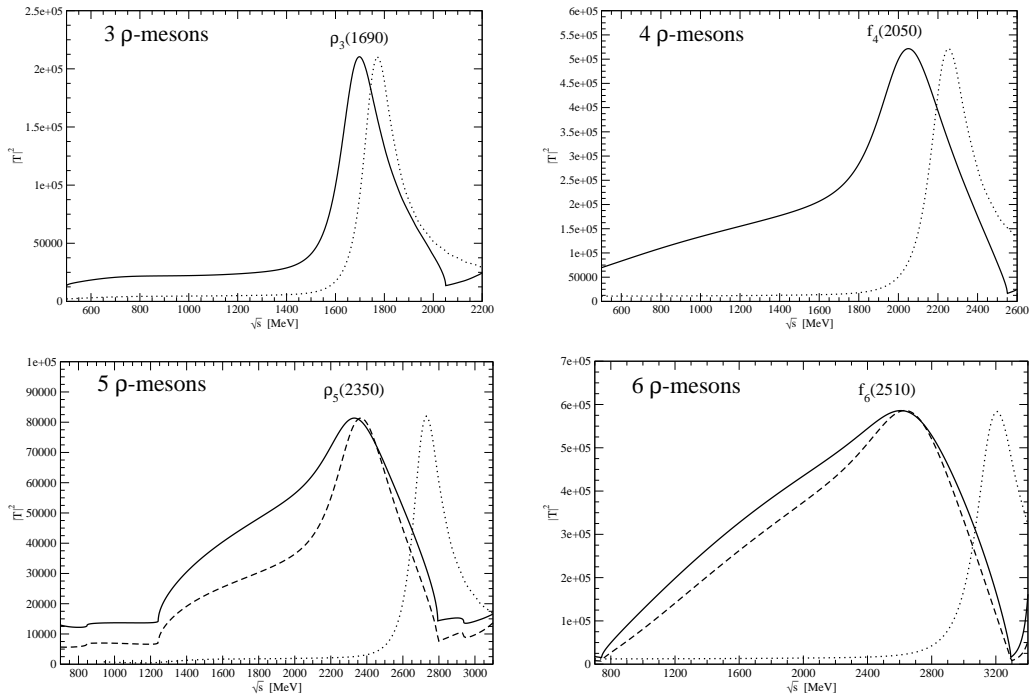
%[htp]
\begin{center}
\makebox[0pt]{\includegraphics[width=.35\linewidth]{figure5a.eps}\hspace{0.5cm}
\includegraphics[width=.35\linewidth]{figure5b.eps}}\\\vspace{0.3cm}
\makebox[0pt]{\includegraphics[width=.35\linewidth]{figure5c.eps}\hspace{0.5cm}
\includegraphics[width=.35\linewidth]{figure5d.eps}}
   \caption{Modulus squared of the unitarized multi-$\rho$ amplitudes.
   dotted line: only single-scattering. Solid lines correspond to the prediction of the model. Dashed lines come from making a small change in a cut off. The solid one is the one used. 
 (The dashed and dotted lines have been normalized to
   the peak of the solid line for the sake of comparison 
   of the position 
   of the maxima)}
     \label{fig:T2s}
\end{center}
\end{figure}

In fig.~\ref{fig:T2s} we show the modulus squared of the amplitudes for
different number of $\rho$ mesons considering only the single scattering
mechanisms (dotted line) and the full model (solid and dashed lines). The
dotted and dashed curves have been normalized to the peaks of the
corresponding full result for the sake of comparison of the position of
the maximum.
The difference between the dashed and solid lines can be
considered as an
estimate of the error but the variation in the position of the maximum
is small.

We clearly see that the amplitudes show  
pronounced bumps which we associate to the resonances 
labeled in the figures.
The position of  
the maxima can be associated to the masses of the corresponding
resonances. The widths can be induced from the figure and agree with the experimental data.

\begin{figure}[!t]
\begin{center}
\includegraphics[width=0.50\textwidth]{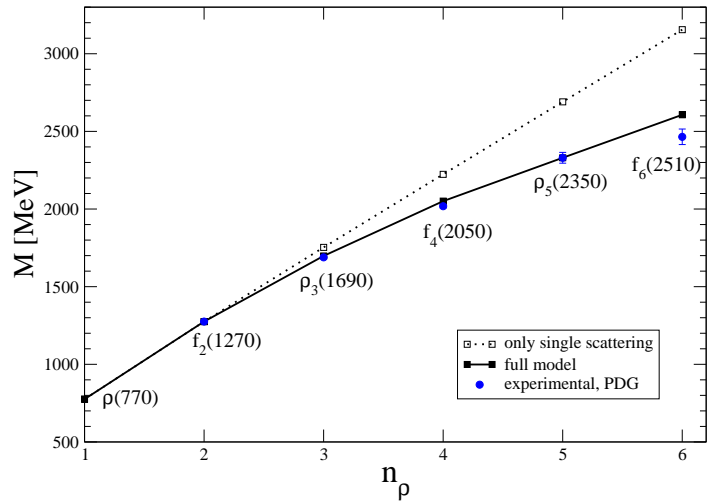}
\caption{Masses of the dynamically generated states as a function of the
number of constituent $\rho(770)$ mesons, $n_\rho$. Only single
scattering contribution (dotted line); full model (solid line);
experimental values from the PDG, (circles).
}
\label{fig:Mvsn}
\end{center}
\end{figure}

 The masses obtained are compared with those in the PDG in fig. \ref{fig:Mvsn}. As we can see, the agreement obtained is excellent. Granted that Nature is always more subtle that any picture that we can make of it, the agreement found for the different states is certainly impressive, with no free parameters used. This is certainly a result worth thinking about which should encourage further searches in this direction with likely interesting surprises ahead. 
In this respect of work the work on multirho states has been extended in \cite{junkomulti} to study systems with a $K^*$ and many $\rho$ states with their spins aligned. Neat peaks are obtained at positions that correspond to the known states $K^*_2(1430)$, $K^*_3(1780)$, $K^*_4(2045)$, $K^*_5(2380)$ and a not yet discovered $K_6^*$ resonance is predicted. 

\section{Conclusions}
   
   The finding about the cancellations of the off shell two body amplitudes with three body terms stemming from the same chiral Lagrangians is most welcome feature which renders the Faddeev equations unambiguous. The results obtained with two meson one baryon systems, reproducing the low lying $1/2^+$ baryon states, are surprising and challenge the conventional wisdom about baryons being three quark states.  

Watch for multihadron states, they were
always there but only now they are beginning to be identified. Facilities and collaborations like  BELLE, COMPASS, WASA/COSY, etc., will have something to say about this topic in the future.


\begin{thebibliography}{99}
\itemsep -2pt 

%\cite{Faddeev:1960su}
\bibitem{Faddeev}
  L.~D.~Faddeev,
  %``Scattering theory for a three particle system,''
  Sov.\ Phys.\ JETP {\bf 12}, 1014 (1961)
  [Zh.\ Eksp.\ Teor.\ Fiz.\  {\bf 39}, 1459 (1960)].
  %%CITATION = ZETFA,39,1459;%%
  
 %\cite{Alt:1967fx}
\bibitem{Alt:1967fx}
  E.~O.~Alt, P.~Grassberger and W.~Sandhas,
  %``Reduction of the three - particle collision problem to multichannel two -
  %particle Lippmann-Schwinger equations,''
  Nucl.\ Phys.\  B {\bf 2} (1967) 167.
  %%CITATION = NUPHA,B2,167;%%
  
  %\cite{Kaiser:1995eg}
\bibitem{Kaiser:1995eg}
  N.~Kaiser, P.~B.~Siegel and W.~Weise,
  %``Chiral Dynamics And The Low-Energy Kaon - Nucleon Interaction,''
  Nucl.\ Phys.\  A {\bf 594}, 325 (1995).
%  [arXiv:nucl-th/9505043].
  %%CITATION = NUPHA,A594,325;%%

%\cite{Oset:1997it}
\bibitem{angels}
  E.~Oset and A.~Ramos,
  %``Non perturbative chiral approach to s-wave anti-K N interactions,''
  Nucl.\ Phys.\  A {\bf 635}, 99 (1998).
%  [arXiv:nucl-th/9711022].
  %%CITATION = NUPHA,A635,99;%%

%\cite{Oller:2000fj}
\bibitem{ollerulf}
  J.~A.~Oller and U.~G.~Meissner,
   %``Chiral dynamics in the presence of bound states: Kaon nucleon  %interactions
  %revisited,''
  Phys.\ Lett.\  B {\bf 500}, 263 (2001).
%  [arXiv:hep-ph/0011146].
  %%CITATION = PHLTA,B500,263;%%
  
  %\cite{Jido:2003cb}
\bibitem{Jido:2003cb}
  D.~Jido, J.~A.~Oller, E.~Oset, A.~Ramos, U.~G.~Meissner,
  %``Chiral dynamics of the two Lambda(1405) states,''
  Nucl.\ Phys.\  {\bf A725}, 181-200 (2003).
 % [nucl-th/0303062].

 %\cite{Oset:2001cn}
\bibitem{Oset:2001cn}
  E.~Oset, A.~Ramos and C.~Bennhold,
  %``Low lying S = -1 excited baryons and chiral symmetry,''
  Phys.\ Lett.\  B {\bf 527}, 99 (2002)
  [Erratum-ibid.\  B {\bf 530}, 260 (2002)].
%  [arXiv:nucl-th/0109006].
  %%CITATION = PHLTA,B527,99;%%
 
 %\cite{Hyodo:2002pk}
\bibitem{Hyodo:2002pk}
  T.~Hyodo, S.~I.~Nam, D.~Jido and A.~Hosaka,
   %``Flavor SU(3) breaking effects in the chiral unitary model for meson baryon
  %scatterings,''
  Phys.\ Rev.\  C {\bf 68}, 018201 (2003).
%  [arXiv:nucl-th/0212026].
  %%CITATION = PHRVA,C68,018201;%% 
  
%\cite{Borasoy:2005ie}
\bibitem{Borasoy:2005ie}
  B.~Borasoy, R.~Nissler and W.~Weise,
  %``Chiral dynamics of kaon nucleon interactions, revisited,''
  Eur.\ Phys.\ J.\  A {\bf 25}, 79 (2005).
%  [arXiv:hep-ph/0505239].
  %%CITATION = EPHJA,A25,79;%%
  
  
  %\cite{Oller:2006jw}
\bibitem{Oller:2006jw}
  J.~A.~Oller,
  %``On the strangeness -1 S-wave meson baryon scattering,''
  Eur.\ Phys.\ J.\  A {\bf 28}, 63 (2006).
%  [arXiv:hep-ph/0603134].
  %%CITATION = EPHJA,A28,63;%%



%\cite{Borasoy:2006sr}
\bibitem{Borasoy:2006sr}
  B.~Borasoy, U.~G.~Meissner and R.~Nissler,
  %``K- p scattering length from scattering experiments,''
  Phys.\ Rev.\  C {\bf 74}, 055201 (2006).
%  [arXiv:hep-ph/0606108].
  %%CITATION = PHRVA,C74,055201;%% 
  
  %\cite{Lutz:2001yb}
\bibitem{Lutz:2001yb}
  M.~F.~M.~Lutz and E.~E.~Kolomeitsev,
   %``Relativistic chiral SU(3) symmetry, large N(c) sum rules and meson  baryon
  %scattering,''
  Nucl.\ Phys.\  A {\bf 700}, 193 (2002).
%  [arXiv:nucl-th/0105042].
  %%CITATION = NUPHA,A700,193;%% 
  
   %\cite{Nieves:1999bx}
\bibitem{Nieves:1999bx}
  J.~Nieves and E.~Ruiz Arriola,
  %``Bethe-Salpeter approach for unitarized chiral perturbation theory,''
  Nucl.\ Phys.\  A {\bf 679}, 57 (2000)
  %[arXiv:hep-ph/9907469].
  %%CITATION = NUPHA,A679,57;%% 
    
   %\cite{MartinezTorres:2007sr}
\bibitem{MartinezTorres:2007sr}
  A.~Martinez Torres, K.~P.~Khemchandani, E.~Oset,
  %``Three body resonances in two meson-one baryon systems,''
  Phys.\ Rev.\  {\bf C77}, 042203 (2008).
.

%\cite{MartinezTorres:2008gy}
\bibitem{MartinezTorres:2008gy}
  A.~Martinez Torres, K.~P.~Khemchandani, L.~S.~Geng, M.~Napsuciale, E.~Oset,
  %``The X(2175) as a resonant state of the phi K anti-K system,''
  Phys.\ Rev.\  {\bf D78}, 074031 (2008).
  
 %\cite{Torres:2011jt}
\bibitem{Torres:2011jt}
  A.~Martinez Torres, D.~Jido, Y.~Kanada-En'yo,
  %``Theoretical study of the $KK\bar K$ system and dynamical generation of the K(1460) resonance,''
  Phys.\ Rev.\  {\bf C83}, 065205 (2011).
  [arXiv:1102.1505 [nucl-th]].
 
 
  
 %\cite{Khemchandani:2008rk}
\bibitem{Khemchandani:2008rk}
  K.~P.~Khemchandani, A.~Martinez Torres, E.~Oset,
  %``The N*(1710) as a resonance in the pi pi N system,''
  Eur.\ Phys.\ J.\  {\bf A37}, 233-243 (2008).
 
  
  %\cite{Epelbaum:2000mx}
\bibitem{epelbaum}
  E.~Epelbaum {\it et. al.},
  %``The three- and four-nucleon systems from chiral effective field theory,''
  Phys.\ Rev.\ Lett.\  {\bf 86}, 4787 (2001)


  
%\cite{Oller:1997ti}
\bibitem{npa}
  J.~A.~Oller and E.~Oset,
  %``Chiral symmetry amplitudes in the S-wave isoscalar and isovector  channels
  %and the sigma, f0(980), a0(980) scalar mesons,''
  Nucl.\ Phys.\  A {\bf 620}, 438 (1997)
  [Erratum-ibid.\  A {\bf 652}, 407 (1999)].
  %%CITATION = NUPHA,A620,438;%% 
  
 %\cite{Inoue:2001ip}
\bibitem{Inoue}
  T.~Inoue, E.~Oset and M.~J.~Vicente Vacas,
  %``Chiral unitary approach to S-wave meson baryon scattering in the
  %strangeness S=0 sector,''
  Phys.\ Rev.\  C {\bf 65}, 035204 (2002)
 % [arXiv:hep-ph/0110333].
  %%CITATION = PHRVA,C65,035204;%%

%\cite{Garcia-Recio:2003ks}
\bibitem{Garcia-Recio:2003ks}
  C.~Garcia-Recio, M.~F.~M.~Lutz and J.~Nieves,
  %``Quark mass dependence of s-wave baryon resonances,''
  Phys.\ Lett.\  B {\bf 582}, 49 (2004)
  %[arXiv:nucl-th/0305100].
  %%CITATION = PHLTA,B582,49;%%
  

%\cite{Oller:1998zr}
\bibitem{nsd}
  J.~A.~Oller and E.~Oset,
  %``N/D description of two meson amplitudes and chiral symmetry,''
  Phys.\ Rev.\  D {\bf 60}, 074023 (1999).
  %%CITATION = PHRVA,D60,074023;%%


%\cite{LlanesEstrada:2003us}
\bibitem{felipe}
  F.~J.~Llanes-Estrada, E.~Oset, V.~Mateu,
  %``Is the Theta+ a K pi N bound state?,''
  Phys.\ Rev.\  {\bf C69}, 055203 (2004).
  [nucl-th/0311020].





 
 
 %\cite{MartinezTorres:2007rz}
\bibitem{MartinezTorres:2007rz}
  A.~Martinez Torres, K.~P.~Khemchandani and E.~Oset,
  %``The $\sigma K$ coupling in the chiral unitary approach and the isoscalar
  %$\bar{K}N$, $\bar{K}A$ interaction,''
  Eur.\ Phys.\ J.\  A {\bf 36}, 211 (2008).
  %%CITATION = EPHJA,A36,211;%%

 



 %\cite{MartinezTorres:2008kh}
\bibitem{MartinezTorres:2008kh}
  A.~Martinez Torres, K.~P.~Khemchandani and E.~Oset,
  %``Solution to Faddeev equations with two-body experimental amplitudes as
  %input and application to J^P=1/2^+, S=0 baryon resonances,''
  Phys.\ Rev.\  C {\bf 79}, 065207 (2009)
  %[arXiv:0812.2235 [nucl-th]].
  %%CITATION = PHRVA,C79,065207;%%

  
% 
\bibitem{Jido:2008kp}
  D.~Jido and Y.~Kanada-En'yo,
  %``K anti-K N molecule state with I=1/2 and J^P=1/2^+ studied with three-body
  %calculation,''
  Phys.\ Rev.\  C {\bf 78}, 035203 (2008)
  %[arXiv:0806.3601 [nucl-th]].
  %%CITATION = PHRVA,C78,035203;%%

%\cite{MartinezTorres:2009cw}
\bibitem{conulf}
  A.~Martinez Torres, K.~P.~Khemchandani, U.~G.~Meissner and E.~Oset,
  %``Searching for signatures around 1920-MeV of a N* state of three hadron
  %nature,''
  Eur.\ Phys.\ J.\  A {\bf 41}, 361 (2009)
  %[arXiv:0902.3633 [nucl-th]].
  %%CITATION = EPHJA,A41,361;%%
  
  %\cite{Glander:2003jw}
\bibitem{saphir}
 K.~H.~Glander {\it et al.},
 %``Measurement of gamma p --> K+ Lambda and gamma p --> K+ Sigma0 at photon
 %energies up to 2.6 GeV,''
 Eur.\ Phys.\ J.\  A {\bf 19}, 251 (2004)
 %[arXiv:nucl-ex/0308025].
 %%CITATION = EPHJA,A19,251;%% 

%\cite{Bradford:2005pt}
\bibitem{jefflab}
 R.~Bradford {\it et al.}  [CLAS Collaboration],
 %``Differential cross sections for gamma + p --> K+ + Y for Lambda and  Sigma0
 %hyperons,''
 Phys.\ Rev.\  C {\bf 73}, 035202 (2006)
 %[arXiv:nucl-ex/0509033].
 %%CITATION = PHRVA,C73,035202;%% 

%\cite{Sumihama:2005er}
\bibitem{mizuki}
 M.~Sumihama {\it et al.}  [LEPS Collaboration],
 %``The gamma(pol.) p --> K+ Lambda and gamma(pol.) p --> K+ Sigma0  reactions
 %at forward angles with photon energies from 1.5-GeV to  2.4-GeV,''
 Phys.\ Rev.\  C {\bf 73}, 035214 (2006)
 %[arXiv:hep-ex/0512053].
 %%CITATION = PHRVA,C73,035214;%% 

 
 %\cite{Roca:2010tf}
\bibitem{multirho}
  L.~Roca, E.~Oset,
  %``A description of the f2(1270), rho3(1690), f4(2050), rho5(2350) and f6(2510) resonances as multi-rho(770) states,''
  Phys.\ Rev.\  {\bf D82}, 054013 (2010).
 % [arXiv:1005.0283 [hep-ph]].

%\cite{Molina:2008jw}
\bibitem{Molina:2008jw}
  R.~Molina, D.~Nicmorus and E.~Oset,
  %``The \rho\rho interaction in the hidden gauge formalism and the f_0(1370)
  %and f_2(1270) resonances,''
  Phys.\ Rev.\  D {\bf 78}, 114018 (2008)
 % [arXiv:0809.2233 [hep-ph]].
  %%CITATION = PHRVA,D78,114018;%%
  
  %\cite{Geng:2008gx}
\bibitem{Geng:2008gx}
  L.~S.~Geng and E.~Oset,
  %``Vector meson-vector meson interaction in a hidden gauge unitary approach,''
  Phys.\ Rev.\  D {\bf 79}, 074009 (2009)
  %[arXiv:0812.1199 [hep-ph]].
  %%CITATION = PHRVA,D79,074009;%%


%\cite{Fariborz:2006xq}
\bibitem{Fariborz:2006xq}
  A.~H.~Fariborz,
  %``Mass uncertainties of f0(600) and f0(1370) and their effects on
  %determination of the quark and glueball admixtures of the I = 0 scalar
  %mesons,''
  Phys.\ Rev.\  D {\bf 74}, 054030 (2006)
  %[arXiv:hep-ph/0607105].
  %%CITATION = PHRVA,D74,054030;%%
  
%\cite{Umekawa:2004js}
\bibitem{Umekawa:2004js}
  T.~Umekawa, K.~Naito, M.~Oka and M.~Takizawa,
  %``Light scalar mesons in the improved ladder QCD,''
  Phys.\ Rev.\  C {\bf 70}, 055205 (2004)
  %[arXiv:hep-ph/0403032].
  %%CITATION = PHRVA,C70,055205;%%

%\cite{Anisovich:2001zp}
\bibitem{Anisovich:2001zp}
  A.~V.~Anisovich, V.~V.~Anisovich and V.~A.~Nikonov,
  %``Radiative decays of basic scalar, vector and tensor mesons and the
  %determination of the P-wave q anti-q multiplet,''
  Eur.\ Phys.\ J.\  A {\bf 12}, 103 (2001)
  %[arXiv:hep-ph/0108186].
  %%CITATION = EPHJA,A12,103;%%
  

  %\cite{Nagahiro:2008um}
\bibitem{junko}
  H.~Nagahiro, J.~Yamagata-Sekihara, E.~Oset, S.~Hirenzaki and R.~Molina,
  %``The $\gamma \gamma$ decay of the $f_0(1370)$ and $f_2(1270)$ resonances in
  %the hidden gauge formalism,''
  Phys.\ Rev.\  D {\bf 79}, 114023 (2009)
  %[arXiv:0809.3717 [hep-ph]].
  %%CITATION = PHRVA,D79,114023;%%

  %\cite{MartinezTorres:2009uk}
\bibitem{daizou}
  A.~Martinez Torres, L.~S.~Geng, L.~R.~Dai, B.~X.~Sun, E.~Oset and B.~S.~Zou,
  %``Study of the $J/\psi \to \phi (\omega) f_2(1270)$, $J/\psi \to \phi
  %(\omega) f'_2(1525)$ and $J/\psi \to K^{*0}(892) \bar{K}^{* 0}_2(1430)$
  %decays,''
  Phys.\ Lett.\  B {\bf 680}, 310 (2009).
%  [arXiv:0906.2963 [nucl-th]].
  %%CITATION = PHLTA,B680,310;%%
  
 %\cite{Geng:2009iw}
\bibitem{hanhart}
  L.~S.~Geng, F.~K.~Guo, C.~Hanhart, R.~Molina, E.~Oset and B.~S.~Zou,
  %``Study of the $f_2(1270)$, $f_2'(1525)$, $f_0(1370)$ and $f_0(1710)$ in the
  %$J/\psi$ radiative decays,''
  Eur.\ Phys.\ J.\  A {\bf 44} (2010) 305.
%  [arXiv:0910.5192 [Unknown]].
  %%CITATION = EPHJA,A44,305;%%
  
  %\cite{YamagataSekihara:2010qk}
\bibitem{junkomulti}
  J.~Yamagata-Sekihara, L.~Roca, E.~Oset,
  %``On the nature of the $K^*_2(1430)$, $K^*_3(1780)$, $K^*_4(2045)$, $K^*_5(2380)$ and $K^*6$ as $K^*$ - multi-$\rho$ states,''
  Phys.\ Rev.\  {\bf D82}, 094017 (2010).
%  [arXiv:1010.0525 [hep-ph]].


\end{thebibliography}
\end{document}